\documentclass{article}

\usepackage{PRIMEarxiv}

\usepackage[utf8]{inputenc}
\usepackage[T1]{fontenc}
\usepackage{hyperref}
\usepackage{url}
\usepackage{booktabs}
\usepackage{amsfonts,amsmath}
\usepackage{nicefrac}
\usepackage{microtype}
\usepackage{fancyhdr}
\usepackage{graphicx}
\usepackage{subcaption} 
\usepackage{float}     
\usepackage[section]{placeins}

\graphicspath{{media/}}

\title{Learning Discrete Successor Transitions in Continuous Attractor Networks: Emergence, Limits, and Topological Constraints}

\author{
Daniel Brownell \\
Independent Researcher \\
Cape Town, South Africa \\
\texttt{daniel.brownell@gmail.com}
}

\begin{document}
\maketitle

\begin{abstract}
Continuous attractor networks (CANs) are a well-established class of models for representing low-dimensional continuous variables such as head direction, spatial position, and phase \cite{amari1977,zhang1996,burak2009,khona2022}. In canonical spatial domains, transitions along the attractor manifold are driven by continuous displacement signals—such as angular velocity—provided by sensorimotor systems external to the CAN itself. When such signals are not explicitly provided as dedicated displacement inputs, it remains unclear whether attractor-based circuits can reliably acquire recurrent dynamics that support stable state transitions, or whether alternative predictive strategies dominate.

In this work, we present an experimental framework for training CANs to perform successor-like transitions between stable attractor states in the absence of externally provided displacement signals. We compare two recurrent topologies—a circular ring and a folded ``snake'' manifold—and systematically vary the temporal regime under which stability is evaluated. We find that, under short evaluation windows, networks consistently converge to impulse-driven associative solutions that achieve high apparent accuracy yet lack persistent attractor dynamics. Only when stability is explicitly enforced over extended free-run periods do genuine attractor-based transition dynamics emerge. This suggests that shortcut solutions are the default outcome of local learning in recurrent networks, while attractor dynamics represent a constrained regime rather than a generic result.

Furthermore, we demonstrate that topology strictly limits the capacity for learned transitions. While the continuous ring topology achieves perfect stability over long horizons ($t=120$), the folded snake topology hits a geometric limit characterized by failure at manifold discontinuities, which neither curriculum learning nor basal ganglia–inspired gating can fully overcome.
\end{abstract}


\section{Introduction}

Continuous attractor networks (CANs) provide a principled mechanism for representing continuous variables through localized ``bump'' activity stabilized by recurrent excitation and inhibition \cite{amari1977}. Such networks have been studied extensively as models of head direction cells \cite{zhang1996}, grid cells \cite{burak2009}, and eye position integrators \cite{seung1996}. A defining property of CANs is the existence of a low-dimensional manifold of approximately equivalent stable states.

In canonical spatial domains, transitions along this manifold are driven by continuous displacement signals provided by external sensorimotor systems \cite{skaggs1995,mcnaughton2006}. Under these conditions, the CAN functions primarily as a state integrator. However, many cognitive operations require consistent transitions along an internal representational manifold, even when no explicit displacement signal is provided to drive those transitions.

This distinction raises a fundamental question: to what extent can local, attractor-based circuits learn not only stable state representations but also reliable state transitions in the absence of externally provided displacement signals? Related proposals argue that grid-cell-like representations and displacement-style operations may generalize beyond spatial domains, potentially providing a cortical substrate for internal simulation \cite{hawkins2019,hawkins2021}. 
While successor representations have been extensively studied in reinforcement learning \cite{dayan1993,stachenfeld2017,gershman2018}, these approaches typically abstract away the underlying neural dynamics. Conversely, much of the CAN literature assumes that transition dynamics are hard-coded or strictly sensory-driven \cite{fung2008}.

Crucially, we argue that predictive success does not imply attractor dynamics. We investigate the conditions under which transient predictive activity can be distinguished from genuine stable displacements, identifying a pervasive tendency for networks to learn ``shortcut'' solutions rather than robust attractors \cite{geirhos2020}.

For clarity, learning in all experiments refers to experience-driven modification of recurrent connectivity under externally defined training signals, rather than autonomous planning or action selection.

\section{Model Overview}

\subsection{Continuous Attractor Dynamics}
Each model consists of a recurrent neural population whose activity evolves according to leaky integrator dynamics with lateral excitation and inhibition, consistent with classical neural field formulations \cite{amari1977,fung2008}. Stable attractor states correspond to localized bumps of activity on a low-dimensional manifold. We consider two manifolds:
\begin{itemize}
\item \textbf{Ring topology}: A circular manifold with wraparound connectivity, providing a continuous path for displacement \cite{zhang1996}.
\item \textbf{Snake topology}: A folded linear manifold embedded in two dimensions, introducing boundary effects and anisotropic transition geometry.
\end{itemize}

\begin{figure}[ht!] 
    \centering
    \begin{subfigure}{0.38\textwidth}
        \centering
        \includegraphics[width=\linewidth]{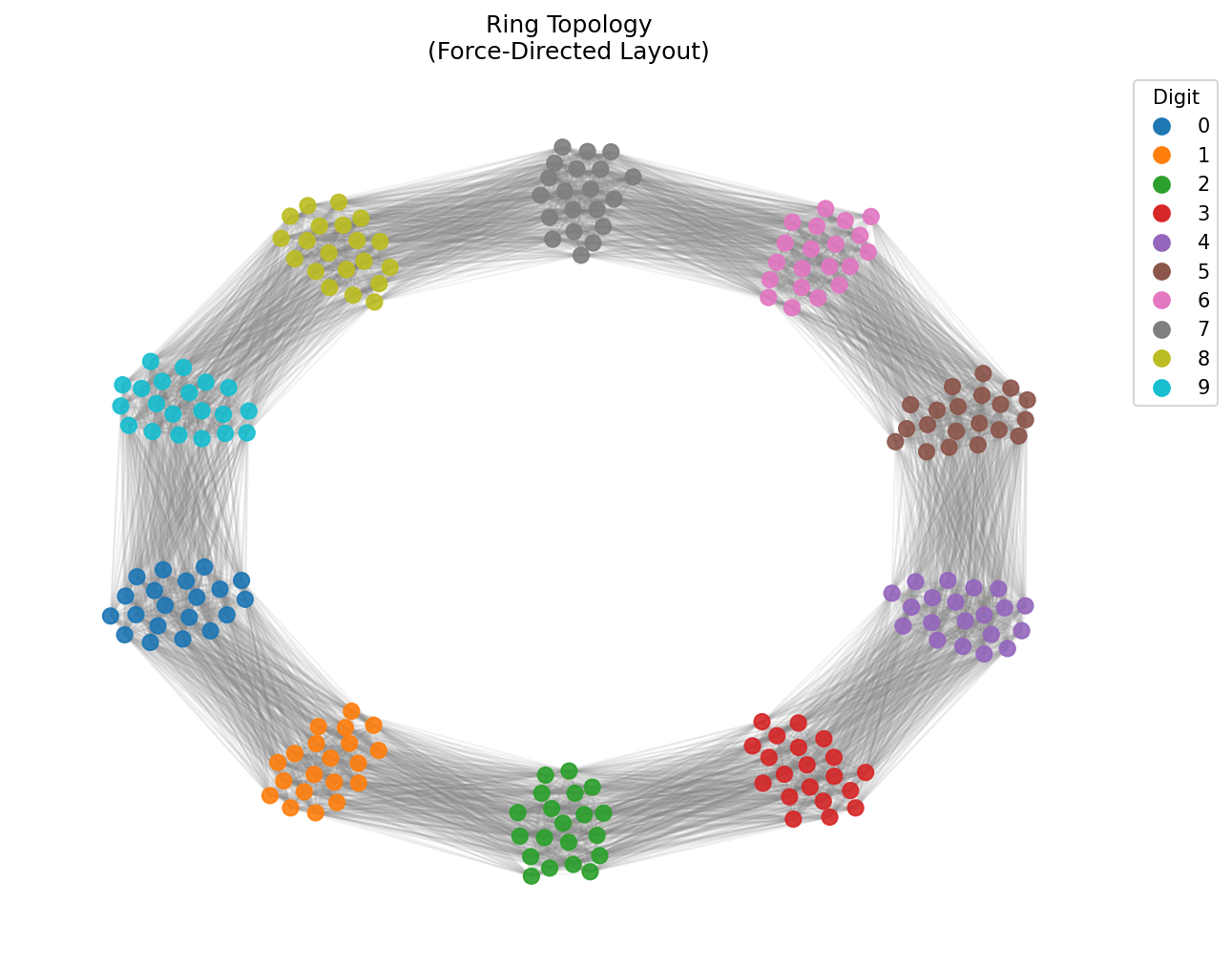}
        \caption{Ring Graph}
        \label{fig:graph_ring}
    \end{subfigure}
    \hfill
    \begin{subfigure}{0.38\textwidth}
        \centering
        \includegraphics[width=\linewidth]{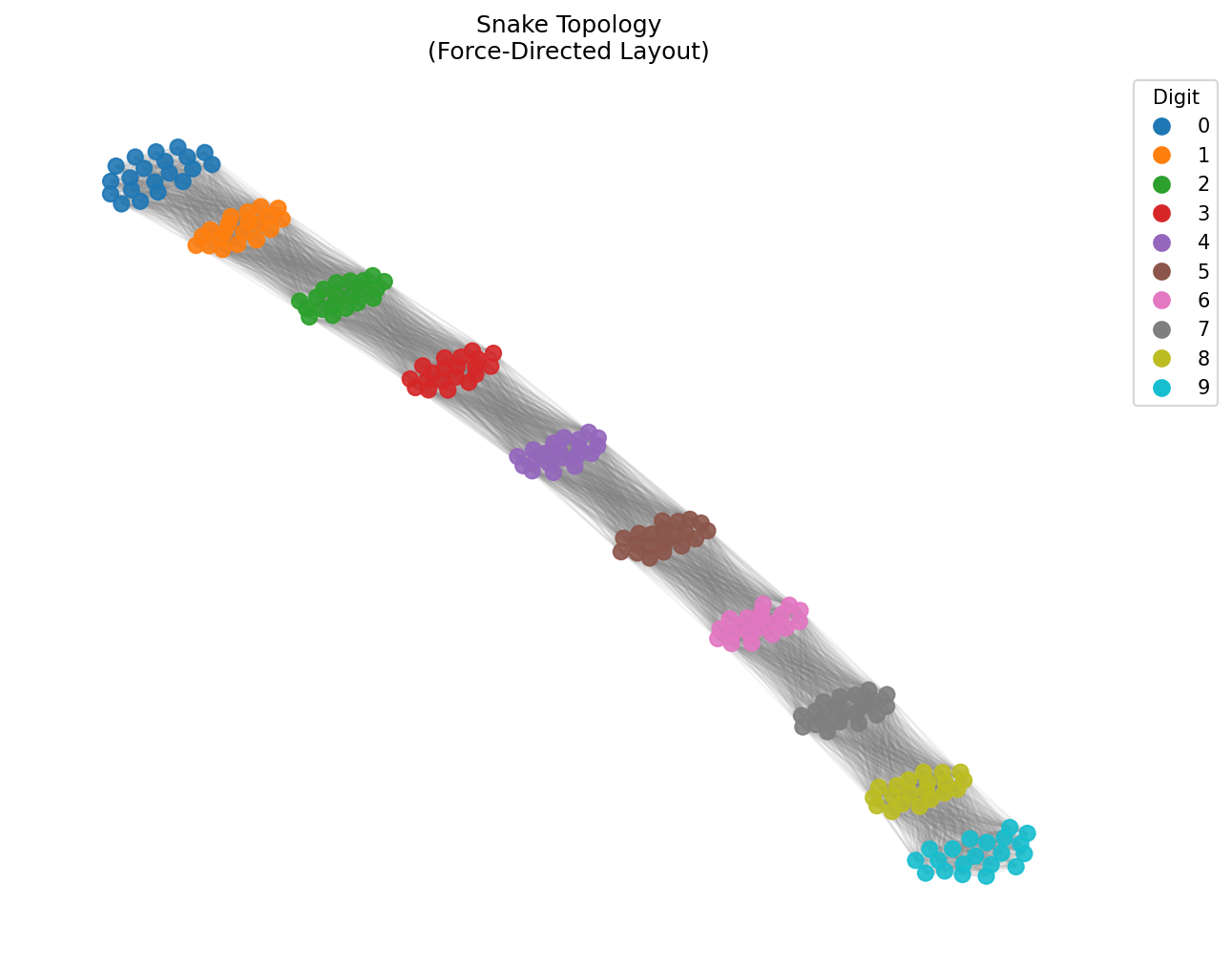}
        \caption{Snake Graph}
        \label{fig:graph_snake}
    \end{subfigure}
    
    \vspace{0.5em}
    
    \begin{subfigure}{0.38\textwidth}
        \centering
        \includegraphics[width=\linewidth]{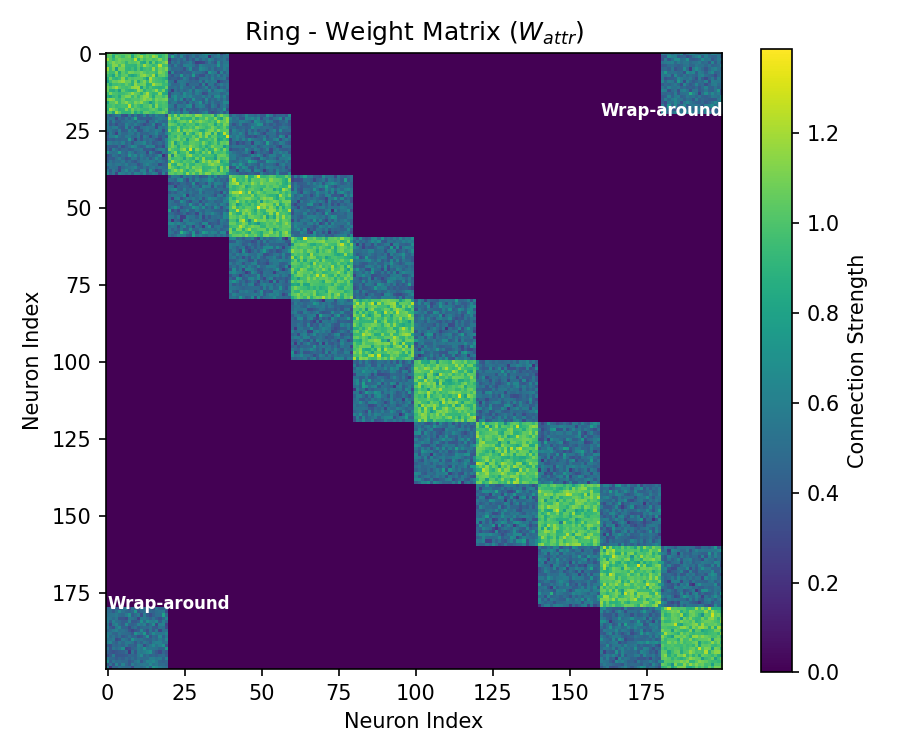}
        \caption{Ring Weights}
        \label{fig:matrix_ring}
    \end{subfigure}
    \hfill
    \begin{subfigure}{0.38\textwidth}
        \centering
        \includegraphics[width=\linewidth]{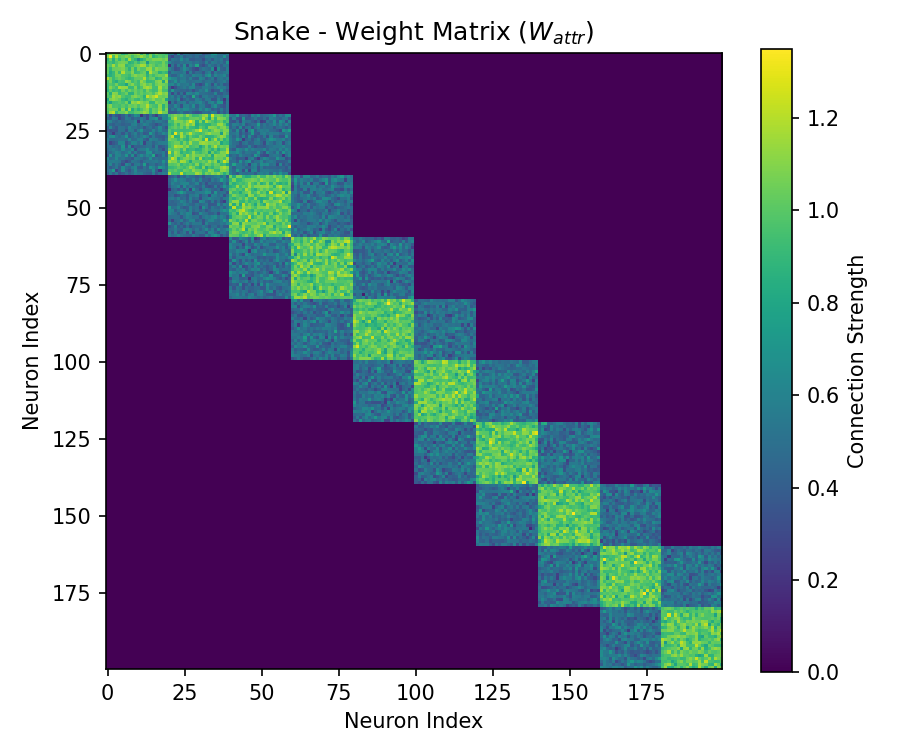}
        \caption{Snake Weights}
        \label{fig:matrix_snake}
    \end{subfigure}
    
    \caption{\textbf{Topology Comparisons.} Force-directed graphs (top) and weight matrices (bottom) revealing the structural differences between Ring and Snake manifolds. Note the continuous connectivity in the Ring matrix versus the hard boundary cutoffs in the Snake matrix.}
    \label{fig:topologies}
\end{figure}

\subsection{Successor Transitions as Learned Displacements}
Successor transitions are implemented as learned recurrent projections that bias the network toward the next attractor state. Functionally, these projections act as learned displacement operators on the attractor manifold, analogous to asymmetric connectivity induced by velocity signals in spatial CANs \cite{burak2009}.

\subsection{Gating and Control}

We evaluate two control policies that regulate when state transitions may occur. In the \emph{Standard} controller, transitions are driven directly by recurrent dynamics without any explicit gating signal. In the \emph{BG-gated} controller, an additional suppressive mechanism—loosely inspired by the basal ganglia—can delay or inhibit transitions, allowing the network additional time to settle before displacement dynamics are engaged \cite{frank2004,oReilly2006,hazy2007}.

Importantly, the BG controller does not introduce new representational structure or alter the underlying attractor dynamics. Rather, it provides an orthogonal control signal that regulates the timing of transitions. Both controllers are evaluated under multiple transition-evaluation regimes described below.

\subsection{Training Curriculum and Evaluation Tasks (Phases A--C)}
\label{sec:curriculum}

\textbf{Curriculum structure.}
Training is organized as a three-phase curriculum in which representational stability is learned before successor transition dynamics are optimized. All experiments share the same underlying recurrent CAN architecture and differ only in topology (Ring vs.\ Snake), controller (Standard vs.\ BG-gated), and the temporal regime used for scoring and selection.

\textbf{Phase A (Digit attractor formation).}
In Phase A the network is trained to represent ten discrete symbolic states (digits 0--9) as stable bump attractors on the manifold. Training adjusts recurrent dynamics so that each digit cue reliably converges to a corresponding stable attractor state. Phase A is considered complete when per-digit readout accuracy reaches 100\% on a validation procedure; we then export a Phase A snapshot (learned recurrent state and prototypes) which is used to initialize downstream training.

\textbf{Phase B (Counting curriculum; optional).}
Phase B is an intermediate curriculum stage intended to shape sequential dynamics on the manifold before training direct successor transitions. In the \emph{countingOn} condition, Phase B trains the network under a structured traversal objective consistent with counting. In the ablation \emph{countingOff}, Phase B is skipped and Phase C begins immediately from the Phase A snapshot. This comparison tests whether an explicit intermediate traversal curriculum changes the optimization path for successor learning, particularly in topologies where local geometry may bias the network toward shortcut solutions.

\textbf{Phase C (Successor transitions).}
Phase C trains the network to implement successor-like transitions (e.g., $x \mapsto x+1$ modulo 10) as learned displacements on the attractor manifold. After an initial digit state is established, the model is required to transition to the successor state and maintain that state autonomously during a free-run period. Phase C admits two solution classes: short-horizon impulse-like transients versus long-horizon self-sustaining attractors.

\textbf{Controllers.}
Both Standard and BG-gated controllers operate on the same underlying recurrent network. The Standard controller allows transitions to proceed immediately under learned recurrent dynamics. The BG-gated controller adds a suppressive timing signal that can delay or inhibit transitions (e.g., waiting steps and optional action sampling), intended to provide additional settling time before displacement dynamics are engaged. In all cases, the controller modulates \emph{when} transitions are permitted rather than altering the Phase A representational space.

\subsection{Automated Hyperparameter Search (Megatune Phase B/C)}
\label{sec:megatune}

\textbf{Motivation.}
The Phase C task admits both (i) shortcut solutions that briefly activate the target state but decay during free-run and (ii) attractor-consistent solutions that sustain the target state over extended horizons. When evaluation is restricted to short temporal windows near transition onset, both solution classes can achieve similarly high prediction accuracy. As a result, naïve training and model selection procedures systematically favor transient solutions that satisfy the loss function without establishing stable attractor dynamics. To make comparisons between topologies and controllers meaningful, we therefore require an automated procedure that explicitly rejects transient solutions during hyperparameter search.

\textbf{Megatuner overview.}
We use an automated multi-stage search procedure (\texttt{megatune-phaseBC}) that tunes parameters governing attractor dynamics (e.g., inhibition strength $k_{\mathrm{inhib}}$, threshold $v_{\mathrm{th}}$, leak/decay $\alpha$, and row-normalized weight magnitudes) and transition-learning parameters (e.g., transition learning rate $\eta_{\mathrm{trans}}$), along with the activity scoring mode used to measure stability (e.g., EMA-based spike activity) and its smoothing parameter. The megatuner is run per topology (Ring and Snake) and per controller family where applicable, producing best-configuration artifacts (e.g., \texttt{megatune\_best\_ring.json}, \texttt{megatune\_best\_snake.json}) that are then used for verification runs and diagnostic plots.

\textbf{Temporal enforcement and windowing.}
Each transition trial is divided into a transition period of length $t_{\mathrm{trans}}$ followed by a free-run tail of length $t_{\mathrm{tail}}$. Candidate configurations are scored over configurable temporal windows (e.g., mean over the whole trial versus late windows that isolate post-transient dynamics). To suppress impulse-based ``cheating,'' attractor-enforced settings (a) exclude the first $k$ time steps after transition onset from evaluation and (b) measure similarity to the target prototype only during the late phase of the trial, after external drive has subsided. This forces candidate solutions to maintain the state representation via recurrent dynamics rather than a brief associative pulse.

\textbf{Composite objective and constraints.}
Candidates are scored with a composite objective trading off (1) successor accuracy and, where applicable, (2) sustained stability during free-run (\emph{sustain}), together with penalties for (3) pathological failure modes such as collapse, silence, or aborted trials. The sustain term is only applied in long-horizon, attractor-enforced regimes (e.g., $t_{\mathrm{trans}}=120$), and is omitted in short-horizon settings where persistence cannot be meaningfully assessed. In addition to weighted scoring, we enforce hard constraints when required, including disallowing impulse-based categorization and requiring that learning and evaluation windows match (preventing optimization on an easier training window than the one used for reporting).

\textbf{Staged search and verification.}
The megatuner uses staged filtering to allocate compute efficiently: an initial guardrail stage rejects obviously unstable candidates with a small number of trials, and later stages evaluate more trials while retaining only a top-$K$ subset. The final selected configuration is then validated by (i) an independent probe step that evaluates candidates under resolved window settings and (ii) full Phase B/C training runs under \emph{countingOn} and \emph{countingOff} conditions. These verification runs support the accuracy summaries, confusion matrices, and transition stability traces reported in Results.

\textbf{Relationship to reported regimes.}
We intentionally contrast megatuner settings corresponding to distinct dynamical regimes. Short-horizon settings (e.g., $t_{\mathrm{trans}}=40$, $t_{\mathrm{tail}}=10$, mean-window scoring, impulse allowed) often yield high apparent accuracy while permitting transient solutions. Long-horizon attractor-enforced settings (e.g., $t_{\mathrm{trans}}=120$, $t_{\mathrm{tail}}=30$, late-window scoring, exclude-first-$k$ steps, impulse disallowed) bias the search toward genuine attractor dynamics and underpin the Ring vs.\ Snake comparison presented here.

\section{Results}

Unless stated otherwise, Phase B/C parameters were selected by the megatuner under the specified temporal regime (short-horizon vs.\ attractor-enforced long-horizon) and then verified by independent probe and full training runs.

\subsection{Associative Shortcut Dynamics as the Default Regime}
We observe that under standard training conditions with short evaluation windows, models consistently achieve high readout accuracy without establishing stable attractors. In this regime, the network learns an impulse-driven association: the initial state triggers a transient spike of activity at the target location, which satisfies the prediction loss but decays rapidly once the input drive is removed.

We term this ``Impulse Cheating'' or Associative Shortcut Dynamics. This is not a rare failure mode but the default outcome of gradient descent on predictive tasks in recurrent networks. The network minimizes energy by predicting the output without investing in the recurrent weights necessary to sustain the activity bump against noise.

\subsection{Regime Selection via Composite Constraints}

Given this tendency, stable attractors require evaluation that rejects transients; increasing the horizon alone (e.g., $t=40 \to 120$) is insufficient without late-window scoring.

Therefore, we define an \textbf{Attractor Regime} that disallows impulse-based categorization, excludes the first $k=8$ steps after transition onset, and scores accuracy only in the late phase after external drive has subsided. This forces candidate solutions to maintain the state via recurrent dynamics rather than a brief associative pulse, aligning with curriculum learning principles \cite{bengio2009}.


\begin{figure}[htbp]
\centering

\captionsetup[subfigure]{
  font=scriptsize,
  labelfont=normalfont,
  position=top,
  skip=1pt,
  justification=centering,
  singlelinecheck=false
}

\newcommand{\tracew}{0.40\linewidth} 
\newcommand{\rowsep}{2pt}             

\subcaptionbox{Standard $t=120$ (Ring)\label{fig:tr_a}}[ \tracew ]{%
  \includegraphics[width=\linewidth]{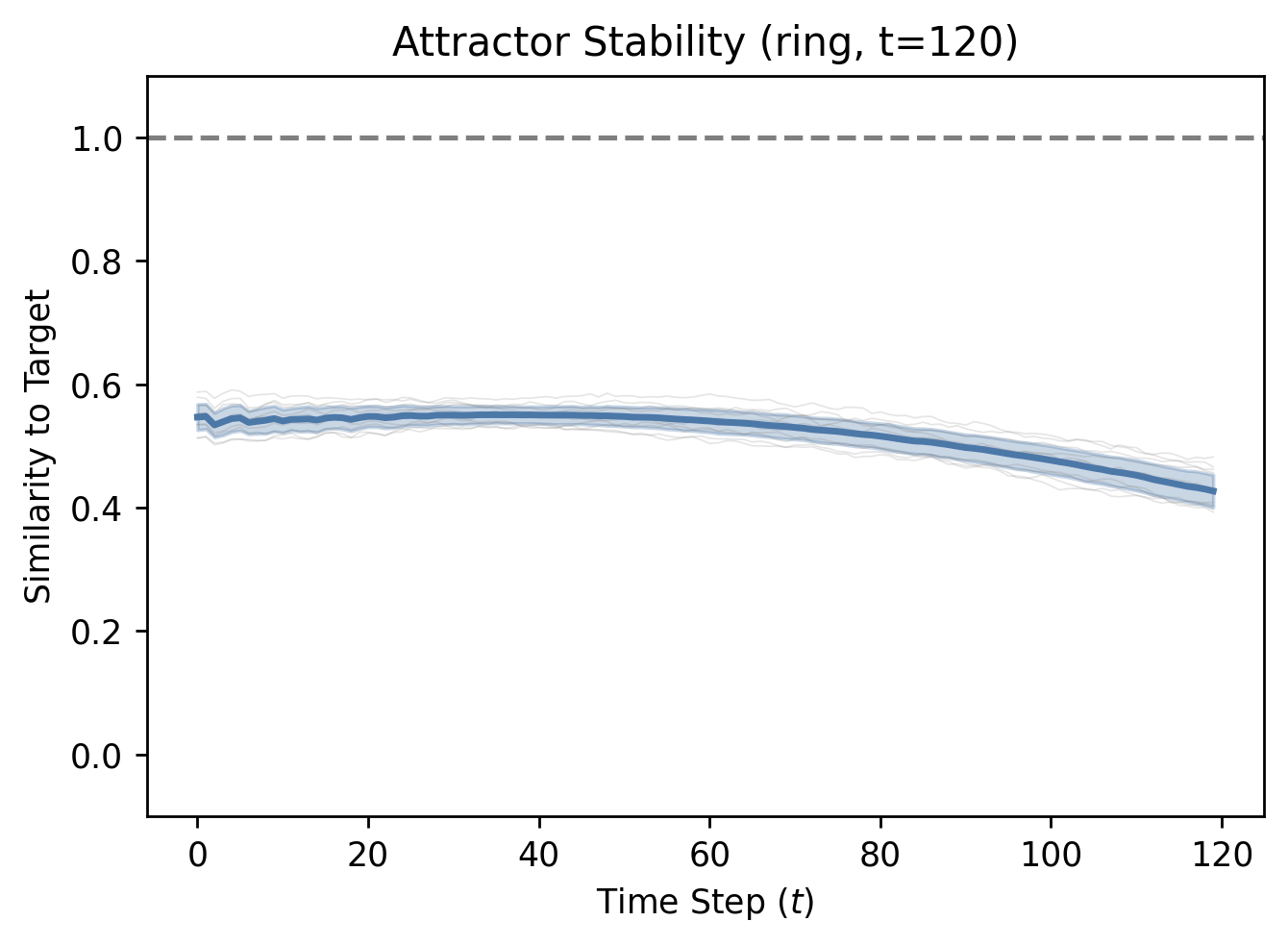}}%
\hfill
\subcaptionbox{Standard $t=120$ (Snake)\label{fig:tr_b}}[ \tracew ]{%
  \includegraphics[width=\linewidth]{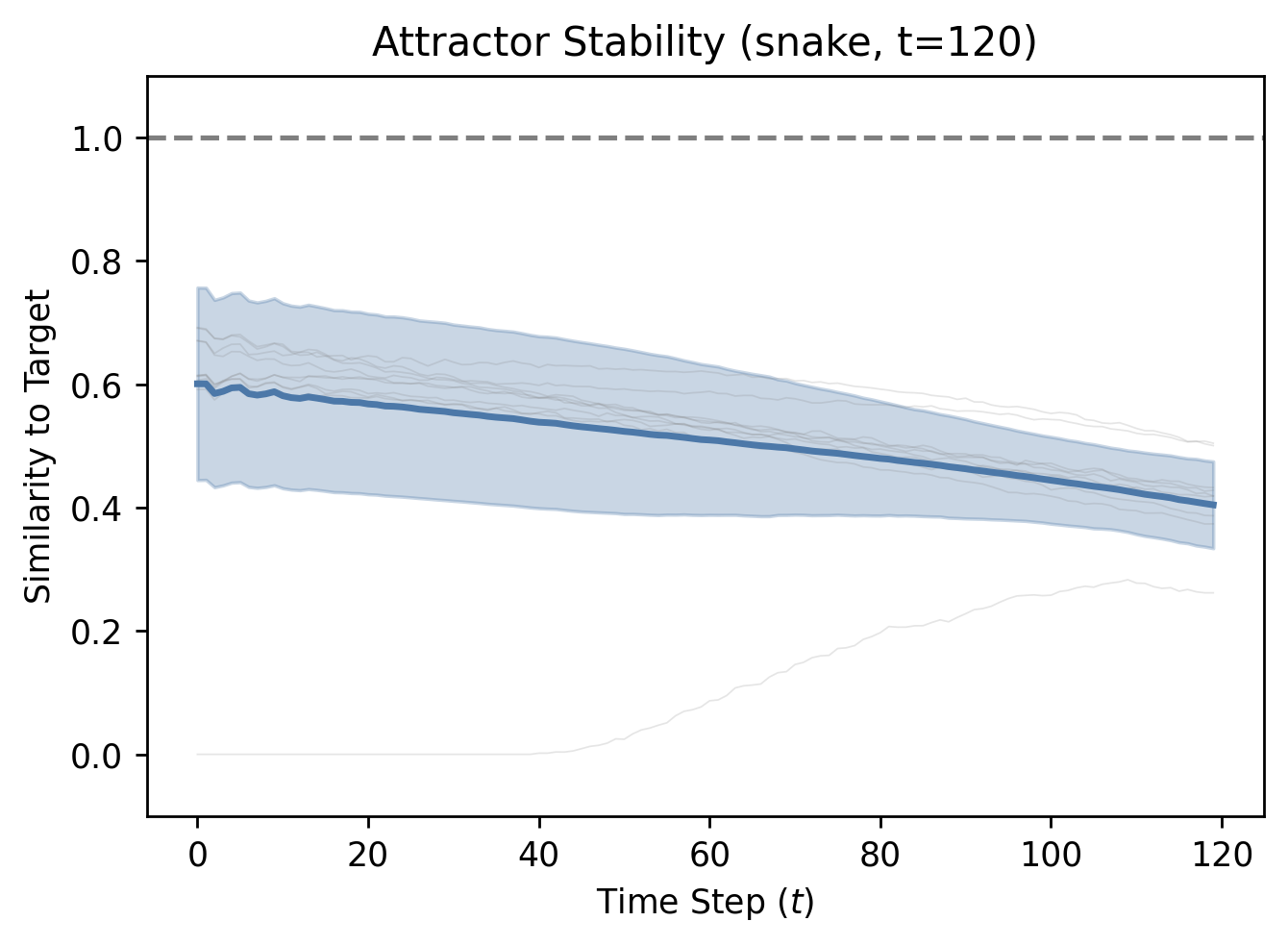}}%

\vspace{\rowsep}

\subcaptionbox{BG-gated $t=120$ (Ring)\label{fig:tr_c}}[ \tracew ]{%
  \includegraphics[width=\linewidth]{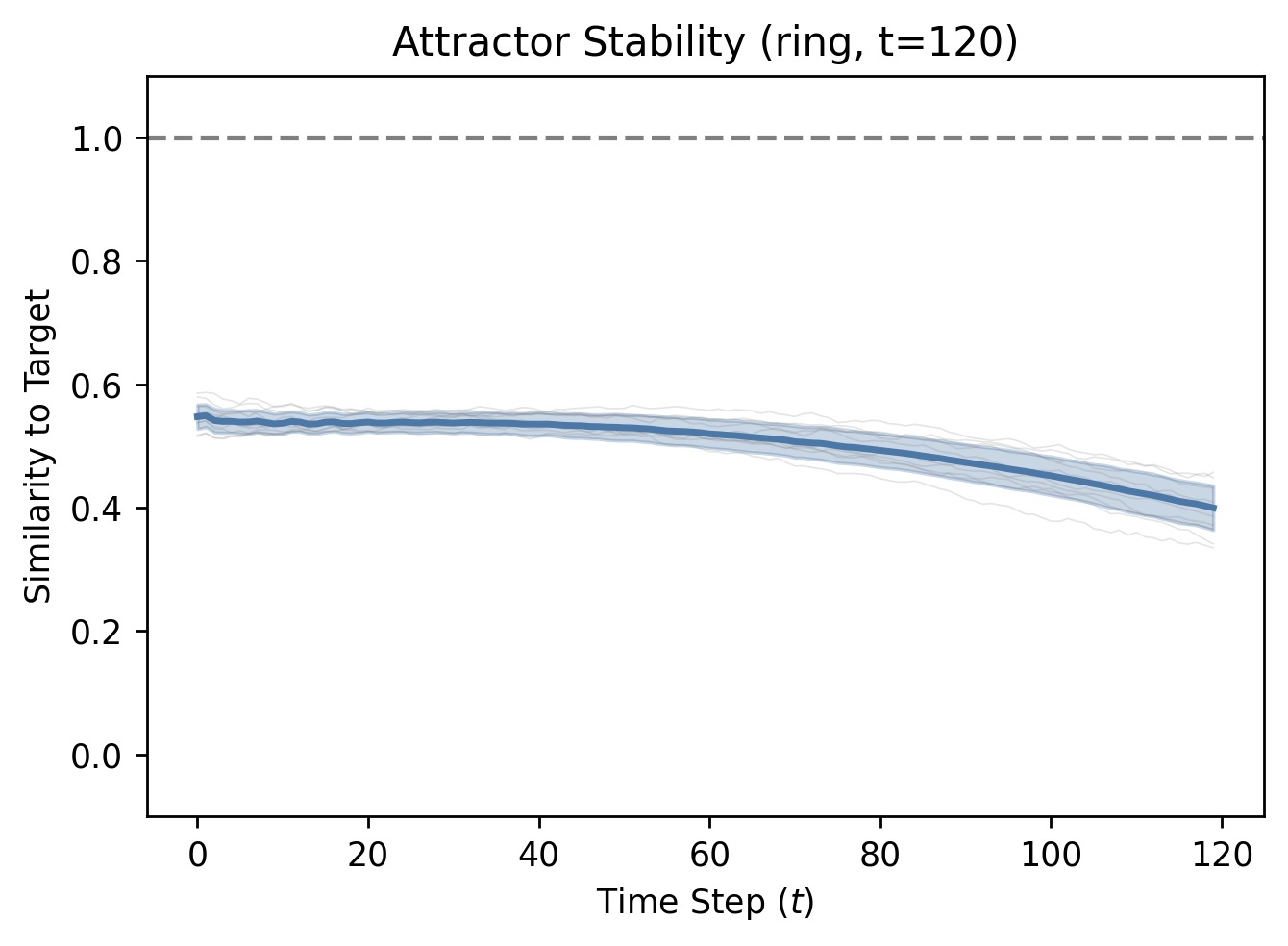}}%
\hfill
\subcaptionbox{BG-gated $t=120$ (Snake)\label{fig:tr_d}}[ \tracew ]{%
  \includegraphics[width=\linewidth]{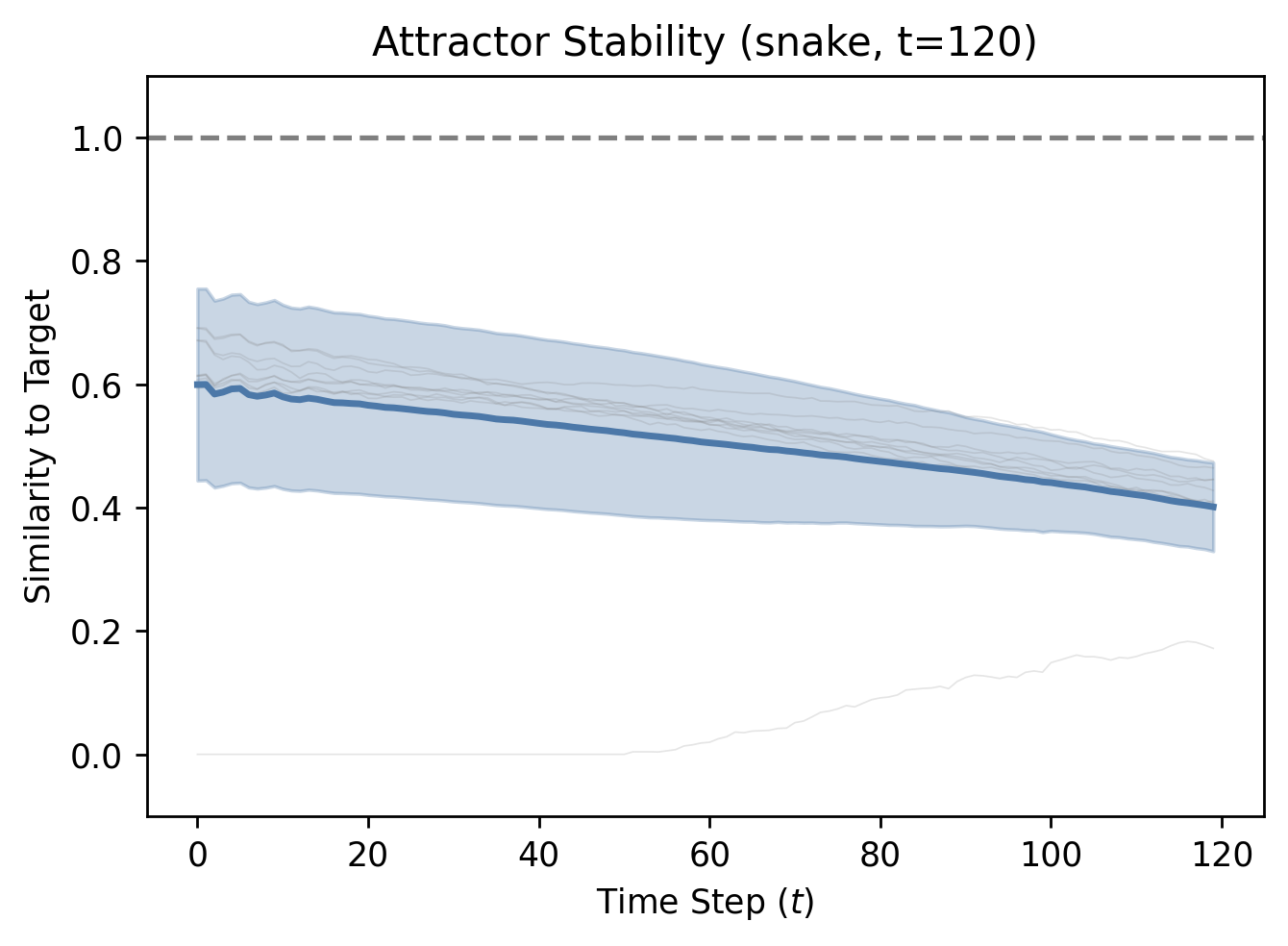}}%

\vspace{\rowsep}

\subcaptionbox{Standard $t=40$ (Ring)\label{fig:tr_e}}[ \tracew ]{%
  \includegraphics[width=\linewidth]{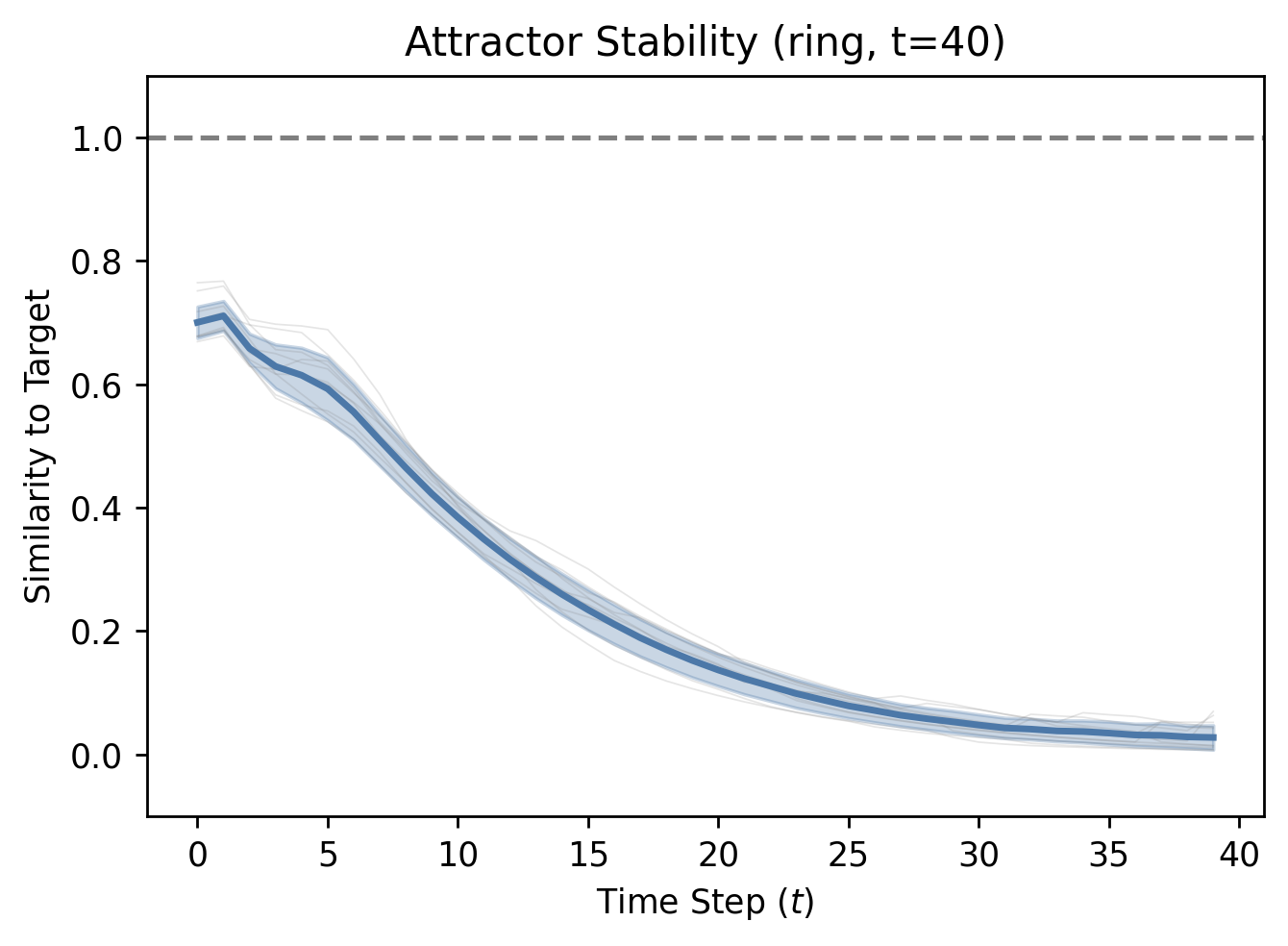}}%
\hfill
\subcaptionbox{Standard $t=40$ (Snake)\label{fig:tr_f}}[ \tracew ]{%
  \includegraphics[width=\linewidth]{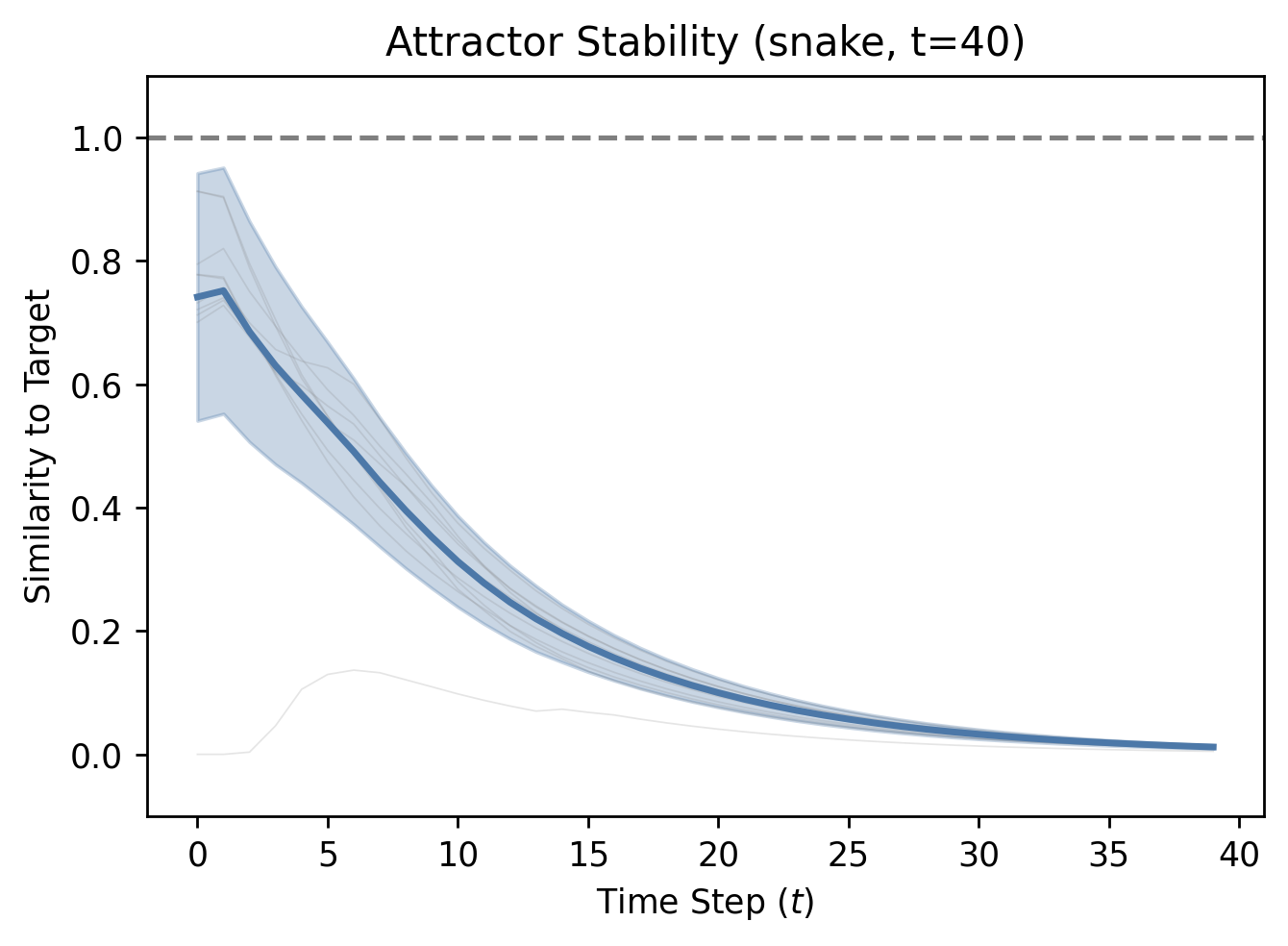}}%

\vspace{\rowsep}

\subcaptionbox{BG-gated $t=40$ (Ring)\label{fig:tr_g}}[ \tracew ]{%
  \includegraphics[width=\linewidth]{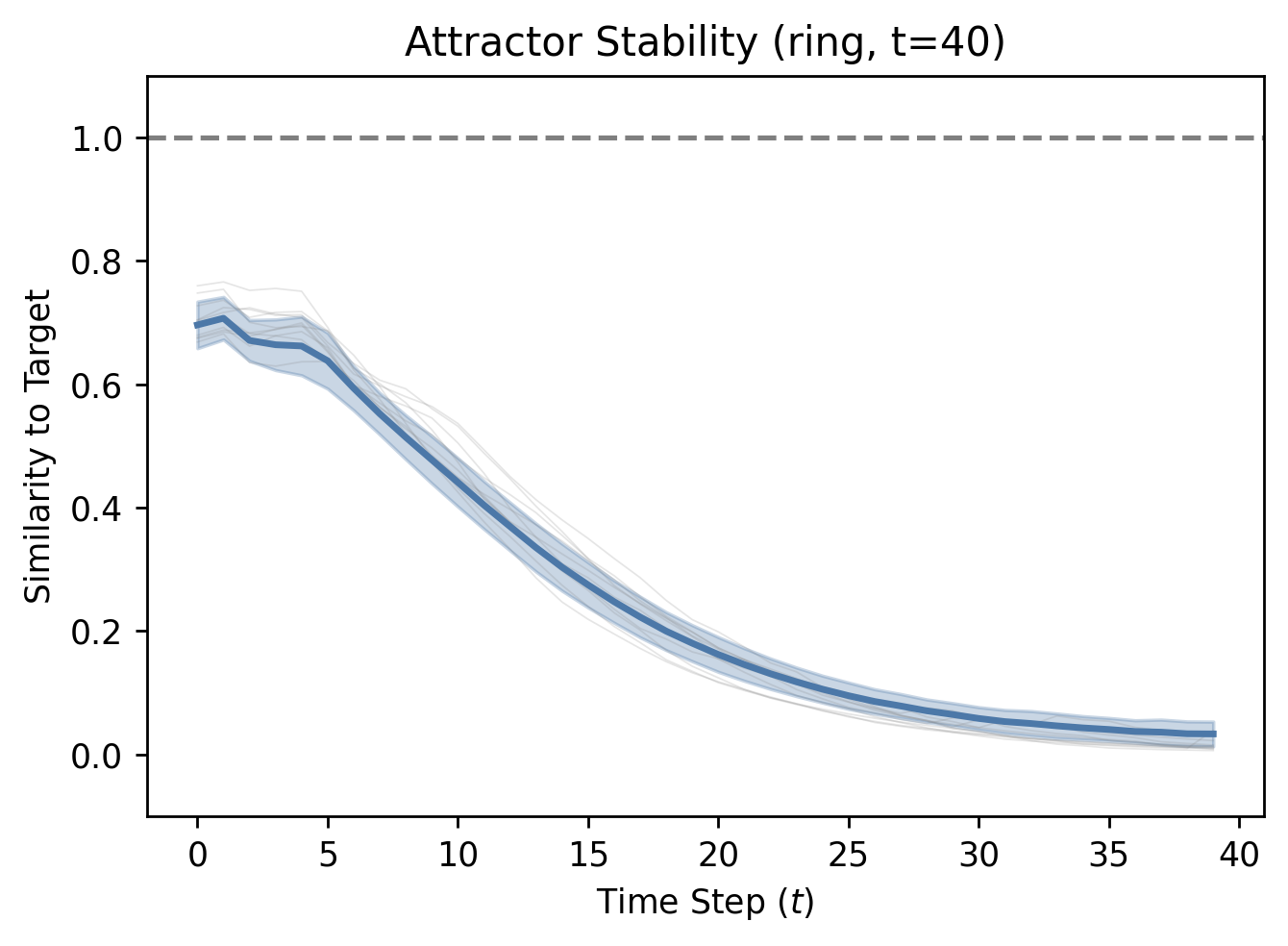}}%
\hfill
\subcaptionbox{BG-gated $t=40$ (Snake)\label{fig:tr_h}}[ \tracew ]{%
  \includegraphics[width=\linewidth]{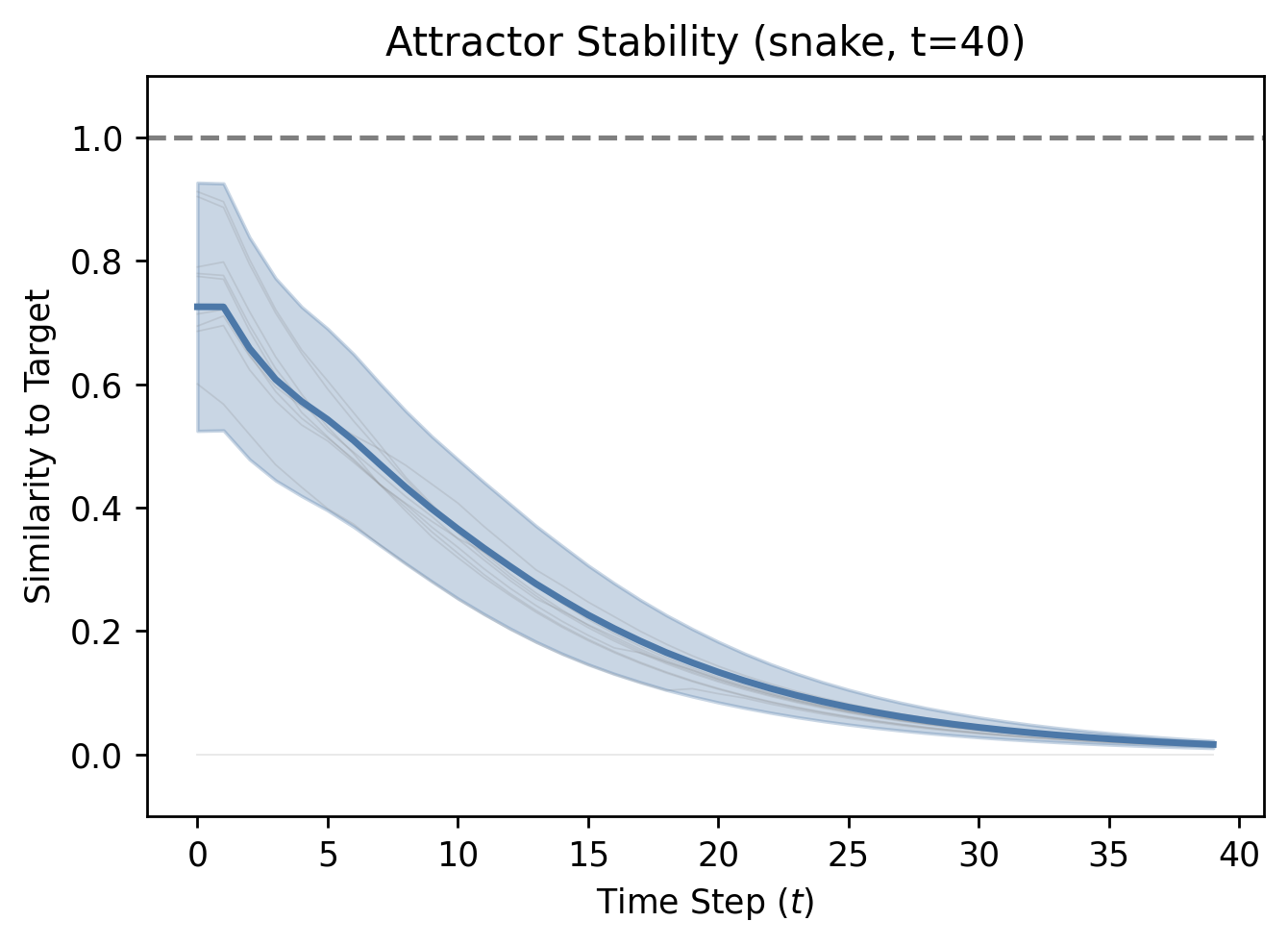}}%

\caption{\textbf{Transition stability traces across horizons and controllers.}
Cosine similarity to the target prototype during free-run dynamics after a successor transition.
Short-horizon evaluation ($t=40$) permits impulse-dominated transients, while long-horizon evaluation ($t=120$)
reveals robust Ring stability and persistent Snake drift that BG gating does not eliminate.}
\label{fig:traces_all}
\end{figure}
\FloatBarrier

\subsection{The Geometric Limit}

The failure of the Snake topology reflects a structural constraint rather than an optimization or control failure. Crucially, this limit only becomes visible once transient shortcut solutions are excluded. Under short-horizon evaluation regimes ($t_{\mathrm{trans}}=40$), the Snake topology achieves near-perfect Phase C accuracy by exploiting impulse-driven dynamics that briefly activate the correct successor state without sustaining it. In contrast, under attractor-enforced long-horizon regimes ($t_{\mathrm{trans}}=120$), these shortcut solutions are no longer sufficient, revealing a systematic geometric failure.

\begin{figure}[H]
    \centering

    \begin{subfigure}{0.40\textwidth}
        \centering
        \includegraphics[width=\linewidth]{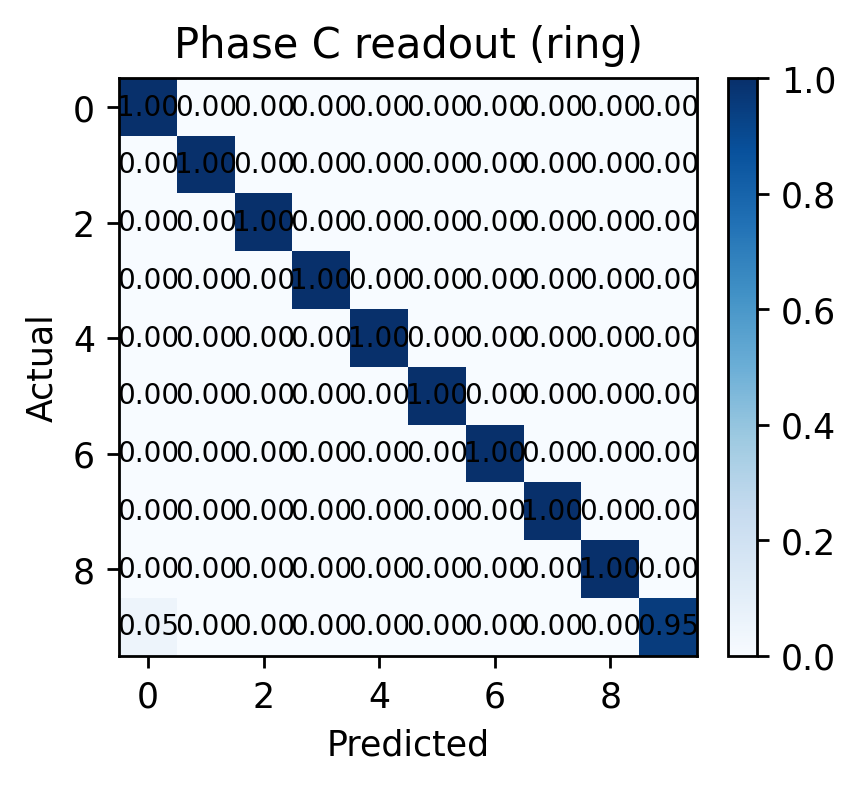}
        \caption{Ring ($t=40$)}
    \end{subfigure}
    \hfill
    \begin{subfigure}{0.40\textwidth}
        \centering
        \includegraphics[width=\linewidth]{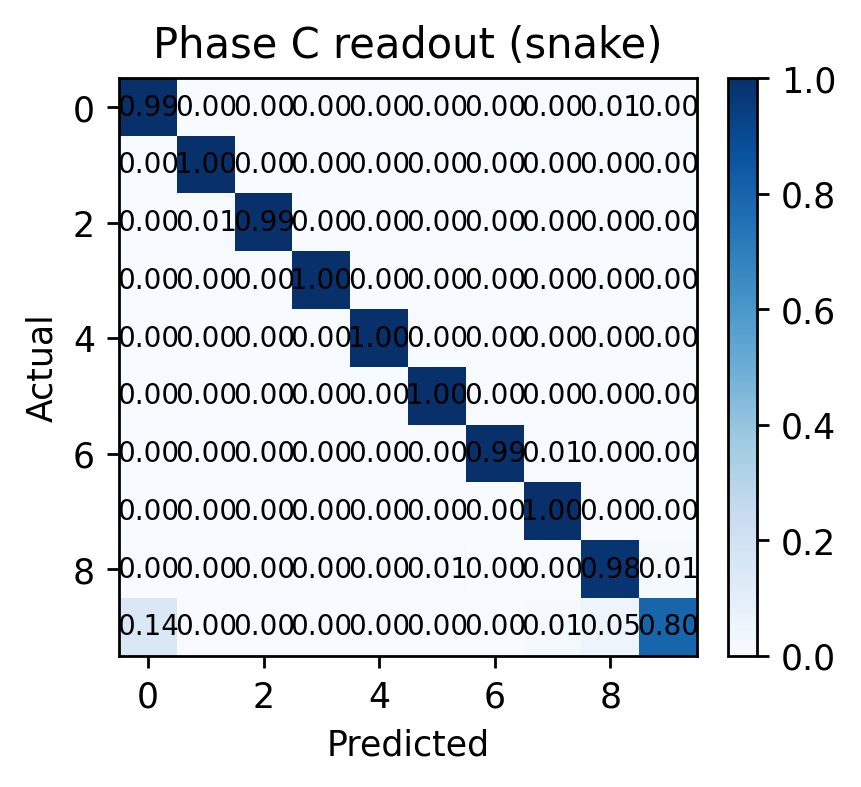}
        \caption{Snake ($t=40$, impulse-permissive)}
    \end{subfigure}

    \vspace{0.5em}

    \begin{subfigure}{0.40\textwidth}
        \centering
        \includegraphics[width=\linewidth]{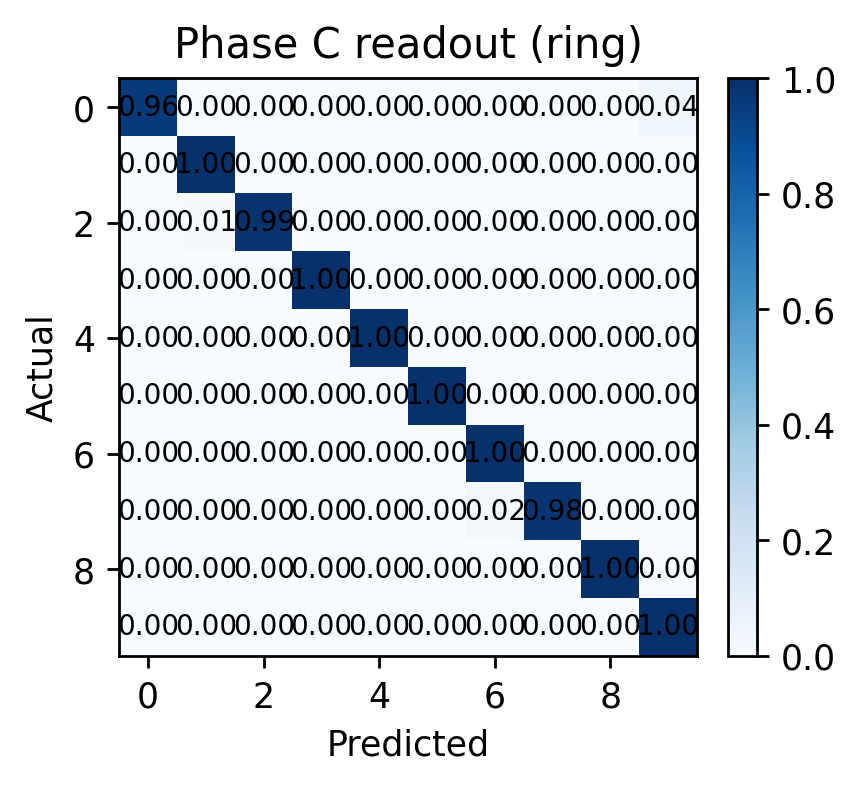}
        \caption{Ring ($t=120$, attractor-enforced)}
    \end{subfigure}
    \hfill
    \begin{subfigure}{0.40\textwidth}
        \centering
        \includegraphics[width=\linewidth]{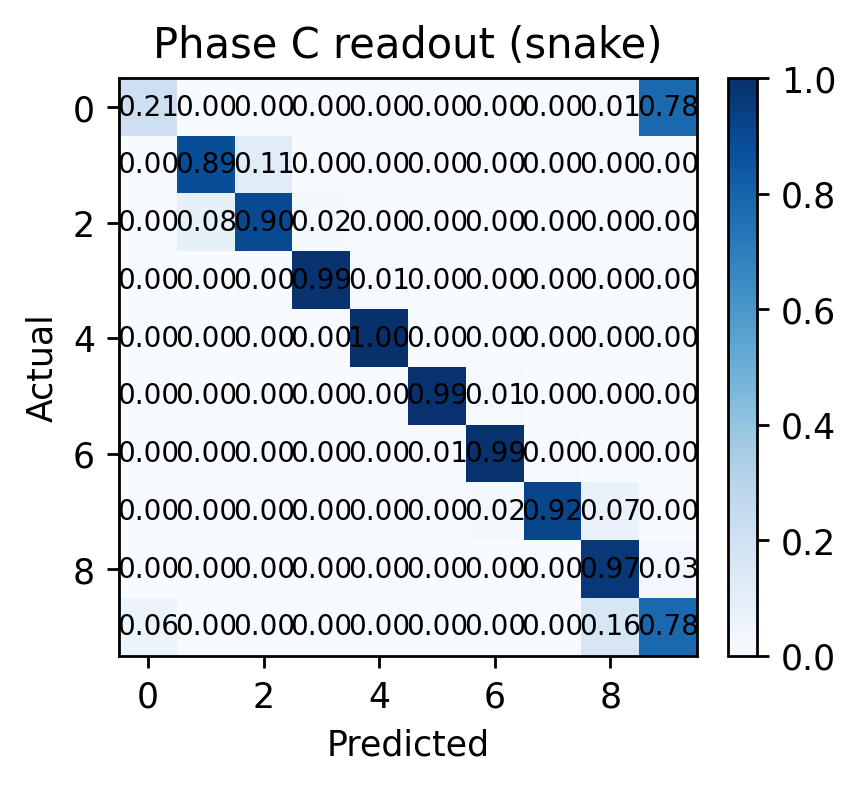}
        \caption{Snake ($t=120$, geometric failure)}
    \end{subfigure}

    \caption{\textbf{Impulse success versus geometric failure in Phase C.}
    Under short-horizon evaluation ($t=40$), both Ring and Snake achieve high apparent accuracy, despite the Snake relying on transient impulse dynamics. Under long-horizon attractor-enforced evaluation ($t=120$), the Ring remains perfectly stable, while the Snake exhibits systematic errors at topological discontinuities (notably $9 \to 0$), revealing a structural geometric limit.}
    \label{fig:confusion_main}
\end{figure}

Figure \ref{fig:confusion_main} compares Phase C confusion matrices for the Standard controller under short-horizon ($t=40$) and long-horizon ($t=120$) regimes. At $t=40$, both Ring and Snake appear successful: the Snake confusion matrix is nearly diagonal, despite relying on transient activity that decays during free-run. At $t=120$, however, the Snake exhibits persistent off-diagonal errors concentrated at manifold discontinuities (most prominently the wrap-around transition $9 \to 0$), while the Ring remains perfectly diagonal.

This divergence demonstrates that the Snake topology supports correct successor \emph{classification} under impulse-cheating regimes but cannot sustain correct successor \emph{dynamics} once stability is enforced. The failure is therefore geometric: the folded manifold introduces non-local discontinuities that cannot be bridged by locally learned recurrent displacements, regardless of curriculum, gating, or extended settling time.

\subsection{Control vs. Geometry}

We compared the Standard controller against a BG-gated controller across both horizons and curricula (Table~\ref{tab:comparison}). 
In the attractor-enforced long-horizon regime ($t=120$), the Snake topology saturates at the same ceiling under both controllers (0.88 with countingOn), indicating a geometric bottleneck rather than insufficient settling time. 
Moreover, BG gating does not reliably improve performance and can degrade it in some settings (notably countingOff), reinforcing that timing control alone cannot compensate for topological discontinuities.

\begin{table}[H]
\centering
\small
\begin{tabular}{lllccl}
\toprule
\textbf{Topology} & \textbf{Controller} & \textbf{Curriculum} & \textbf{Horizon} & \textbf{Acc.} & \textbf{Regime} \\
\midrule
Ring  & Standard & countingOn  & $t=40$  & 1.000 & Short-horizon \\
Ring  & Standard & countingOff & $t=40$  & 1.000 & Short-horizon \\
Snake & Standard & countingOn  & $t=40$  & 1.000 & Short-horizon \\
Snake & Standard & countingOff & $t=40$  & 0.945 & Short-horizon \\
\midrule
Ring  & Standard & countingOn  & $t=120$ & 1.000 & Robust attractor \\
Ring  & Standard & countingOff & $t=120$ & 1.000 & Robust attractor \\
Snake & Standard & countingOn  & $t=120$ & 0.880 & Geom.\ limit \\
Snake & Standard & countingOff & $t=120$ & 0.840 & Geom.\ limit \\
\midrule
Ring  & BG-gated & countingOn  & $t=40$  & 1.000 & Short-horizon \\
Ring  & BG-gated & countingOff & $t=40$  & 1.000 & Short-horizon \\
Snake & BG-gated & countingOn  & $t=40$  & 0.960 & Short-horizon \\
Snake & BG-gated & countingOff & $t=40$  & 0.280 & Control fail \\
\midrule
Ring  & BG-gated & countingOn  & $t=120$ & 1.000 & Robust attractor \\
Ring  & BG-gated & countingOff & $t=120$ & 0.365 & Degraded \\
Snake & BG-gated & countingOn  & $t=120$ & 0.880 & Geom.\ limit \\
Snake & BG-gated & countingOff & $t=120$ & 0.720 & Geom.\ limit \\
\bottomrule
\end{tabular}
\caption{%
\textbf{Control vs.\ geometry across controllers, curricula, and horizons (Phase C).}
At $t=40$, accuracy reflects short-window success and does not certify persistent attractor dynamics (\emph{Short-horizon}).
At $t=120$ under attractor-enforced evaluation, Ring exhibits robust persistence, while Snake is limited by manifold discontinuities (\emph{Geom.\ limit}).
BG gating does not rescue Snake and can degrade performance in some settings (\emph{Control fail}, \emph{Degraded}).}
\label{tab:comparison}
\end{table}

\section{Discussion}

This study began with a simple question: to what extent can a local recurrent circuit learn not only stable state representations, but also reliable state transitions when no explicit displacement signal is provided. In spatial CAN models, transitions are typically driven by an external velocity or displacement input; here we asked whether comparable transition dynamics can be acquired from training alone, and how easily one can confuse ``correct outputs'' with ``correct dynamics'' when evaluating such systems.

\paragraph{Relationship to oscillatory-interference displacement mechanisms.}
In many spatial CAN models, the assumption of an externally provided displacement signal is instantiated not only as an explicit velocity input, but via \emph{oscillatory interference} (OI) mechanisms in which velocity-modulated oscillators generate a structured displacement drive that shifts the bump along an attractor manifold \cite{burgess2008,bush2014}. Hybrid OI+CAN accounts emphasize that recurrent attractor dynamics primarily stabilize and denoise a low-dimensional state manifold, while the displacement computation itself is implemented by an auxiliary circuit that injects a direction- and speed-dependent bias \cite{bush2014}. The present work intentionally \emph{excludes} such dedicated displacement mechanisms in order to isolate what local recurrent learning can and cannot acquire when successor-like transitions must be generated endogenously. Our results therefore should be read as constraints on the hypothesis that a simple local CAN module can learn reliable translation dynamics \emph{without} an explicit displacement drive: absent such structure, optimization is strongly biased toward transient shortcut solutions, and topology can impose hard limits on stable learned transitions.

It is worth noting that many accounts of internal simulation implicitly assume that internally generated sensory or sensorimotor activity (e.g., imagined visual or motor signals) provides an effective displacement input, with the CAN primarily serving to stabilize and integrate that evolving state; the present experiments instead isolate the complementary case in which such displacement-like inputs are absent or underspecified.

\textbf{Prediction accuracy can hide non-attractor solutions.}
Across conditions, we found that Phase C admits at least two qualitatively different solution classes. Under short-horizon, impulse-permissive regimes, the network can achieve high readout accuracy by producing a brief associative pulse at the target state. This looks like a successful successor transition when measured near transition onset, but it does not imply that the target state is stably maintained during free-run. The key methodological point is therefore not merely that ``longer trials'' matter, but that \emph{selection criteria must explicitly reject early transients} (e.g., by excluding the first $k$ steps and scoring only late-trial similarity). Without those constraints, naïve training and selection procedures systematically overestimate success by rewarding solutions that satisfy the loss function transiently rather than sustaining the representational state.

\paragraph{Implications for architectural extensions.}
A natural response to the geometric failure observed in the Snake topology is to posit additional non-local mechanisms capable of bridging discontinuities in the representational manifold. In biological models, this role is often attributed to active dendritic processes or other forms of context-dependent, long-range coupling. A concrete instantiation of this idea in the present setting would be to relax strict locality by introducing sparse long-range connectivity (e.g., a small fraction of background edges) that could couple otherwise disconnected regions of the manifold. Such connectivity might partially mitigate fold-induced failures by providing non-local influence paths, but it would also alter the effective attractor geometry and risks reintroducing shortcut dynamics unless carefully constrained.

From this perspective, our results do not argue against the necessity of additional mechanisms; rather, they help clarify what those mechanisms would need to accomplish. The failure modes identified here arise when successor transitions are delegated entirely to local recurrent attractor dynamics. Any architectural extension that successfully rescues these transitions must therefore implement conditional or non-local displacement structure beyond what a simple CAN provides. In this sense, the present results delineate a boundary: they show which transition behaviors cannot be obtained from local attractor recurrence alone, and thereby sharpen the functional role that additional architectural features would have to play.

\textbf{Attractor-consistent transitions emerge, but only in a constrained regime.}
When evaluated under attractor-enforced settings (late-window scoring, impulse disallowed), the Ring topology supports robust successor transitions and persistent bump stability across long horizons. In that regime, the same network and learning rules that happily converge to shortcut solutions can also support genuine attractor-consistent dynamics, suggesting that the difficulty is not an inherent incapacity to form stable transitions, but the ease with which optimization discovers cheaper predictive strategies unless the task is formulated to rule them out.

\textbf{Topology imposes hard limits on learned transitions.}
The Ring versus Snake comparison shows that even when shortcut solutions are excluded, topology strongly constrains what transitions can be learned by local recurrence. The Snake manifold, which preserves local adjacency but introduces discontinuities at fold boundaries, exhibits systematic errors concentrated at those discontinuities once attractor stability is enforced (notably the $9 \to 0$ wrap-around transition). This pattern is consistent with a geometric bottleneck: locally learned displacement-like dynamics can slide activity along continuous neighborhoods, but cannot reliably implement non-local ``jumps'' across a fold without additional structure or inputs. In this sense, the Snake results help separate two failure modes that can otherwise be conflated: transient cheating (an evaluation artifact) versus genuine geometric limitation (a representational constraint).

\textbf{Control mechanisms modulate timing, not geometry.}
We tested a BG-gated controller intended to provide additional settling time before transitions are engaged. In the attractor-enforced regime, BG gating did not raise the Snake ceiling and in several settings degraded performance, suggesting that timing control alone cannot compensate for discontinuities that require non-local transitions. This does not rule out more complex control schemes, but it supports a narrower conclusion: within the class of controllers tested here, the dominant limitation for the Snake topology is geometric rather than temporal.

\textbf{Implications and scope.}
Our results are deliberately modest: they do not claim that cortical circuits cannot realize CAN-like dynamics, nor do they prescribe a particular biological mechanism. Instead, they identify a practical and conceptual hazard: recurrent networks trained for successor-like prediction will often adopt transient predictive strategies that are difficult to distinguish from attractor dynamics unless evaluation is designed to measure persistence. This matters for any theory that leans on CAN-like modules for internally generated trajectories (e.g., imagined transitions, replay, or ``thought experiments'') in the absence of continuous sensory displacement input. If such circuits are to serve as stable internal simulators, then either (i) additional mechanisms must bias learning toward sustained attractors, or (ii) the relevant computations may rely on dynamics that are predictive but not attractor-based. Our framework makes these alternatives empirically testable by separating ``classification success'' from ``dynamical success'' under controlled regimes.

\textbf{Limitations and future work.}
This work focuses on small synthetic tasks (digits and successor modulo 10) and two hand-designed manifolds, and is intended to probe the dynamical limits of learned successor transitions rather than to provide a comprehensive biological model. Several extensions remain open; however, one natural candidate—bidirectional successor learning—was explicitly tested here and did not resolve the observed failure modes.

In addition to unidirectional successor training ($x \mapsto x+1$), we evaluated multiple variants of \emph{bidirectional} displacement learning in which forward ($+1$) and inverse ($-1$) transitions were trained either sequentially or jointly from scratch. These included curricula that trained all forward transitions followed by all inverse transitions, as well as mixed random-walk training in which $+1$ and $-1$ operators were sampled with equal probability during learning. In all cases, when evaluated under attractor-enforced long-horizon regimes, the resulting dynamics exhibited the same qualitative failure modes observed elsewhere: ``sticky'' attractors that resist displacement, persistent off-by-one errors, and strong asymmetries between forward and inverse operators.

Notably, joint bidirectional training did not mitigate the geometric failures of the folded Snake manifold, nor did it yield a stable bidirectional displacement operator even in the Ring topology. Instead, one direction often dominated while the other degraded, indicating that locally learned recurrent updates do not naturally converge to a group-like bidirectional transition structure. These results push against the idea that continuous attractor networks can internally acquire robust bidirectional displacement dynamics from local Hebbian-style learning alone.

This outcome is consistent with the classical interpretation of CANs as state manifolds whose translation dynamics are ordinarily driven by externally provided displacement or velocity signals, rather than being generated endogenously by the attractor circuit itself. While richer learning rules or explicit latent generative models might discover bidirectional structure under different assumptions, within the learning regime studied here bidirectionality does not rescue folded-manifold failures and instead introduces additional dynamical pathologies.

More broadly, future work could test whether the shortcut-versus-attractor dichotomy identified here persists under alternative learning rules, noise models, larger vocabularies, or multi-step transition objectives; whether relaxing strict locality (e.g., via sparse long-range coupling or explicit displacement-like inputs) can bridge topological discontinuities; and whether hierarchical or multi-area architectures can transform non-local jumps into sequences of locally realizable moves. Across all such extensions, the central methodological lesson remains: evaluation must explicitly measure the dynamical property of interest—late-trial stability—rather than relying on early predictive accuracy, which can systematically reward transient shortcut solutions.

\section{Conclusion}

We introduced an experimental framework for training continuous attractor networks to perform successor-like transitions without externally provided displacement signals, and for disentangling predictive success from attractor-consistent dynamics. We found that short-horizon regimes permit impulse-driven shortcut solutions that achieve high apparent accuracy yet fail to sustain activity during free-run. To obtain meaningful comparisons, model selection must explicitly suppress early transients and evaluate late-trial stability.

Under attractor-enforced evaluation, a Ring topology supports robust learned successor transitions over long horizons, while a folded Snake topology exhibits systematic failures at manifold discontinuities, revealing a geometric limit that simple BG-inspired timing control does not overcome. Together, these results suggest that stable learned transitions are possible but not automatic: they emerge only under constrained evaluation regimes, and their achievable accuracy is strongly shaped by manifold geometry.

\section*{Code Availability}

An open-source reference implementation is available at \url{https://github.com/javadan/can-paper}.

\bibliographystyle{unsrt}
\bibliography{references}

\end{document}